\documentclass{aipproc}

\usepackage{epsfig}

\input paperdef

\begin{document}


\thispagestyle{empty}
\setcounter{page}{0}
\def\thefootnote{\fnsymbol{footnote}}
\onecolumn

\begin{flushright}
BNL--HET--00/46\\
hep-ph/0102318\\
\end{flushright}

\vspace{1cm}

\begin{center}

{\Large\sc {\bf Two-loop Calculations in the MSSM with FeynArts}}
\footnote{Updated version of a talk given at the 
ACAT2000, Fermilab, Oct.~2000.}

\vspace{1cm}

{\sc S.~Heinemeyer$^{1\,}$%
\footnote{
email: Sven.Heinemeyer@bnl.gov
}%
}

\vspace*{1cm}

$^1$ HET, Brookhaven Natl.\ Lab., Upton NY 11973, USA

\end{center}

\vspace*{1cm}

\BC
{\bf\large Abstract}
\EC
Recent electroweak \twol\ corrections in the Minimal Supersymmetric
Standard Model (MSSM) are reviewed. They have been obtained with the help of
the programs \fa\ and \tc, making use of the recently completed MSSM
model file for \fa. Short examples of how to use the two codes
together with the analytic result for the \order{\gf^2\mt^4}
corrections to the $\rho$-parameter in the MSSM are presented.

\def\thefootnote{\arabic{footnote}}
\setcounter{footnote}{0}

\newpage


\twocolumn
\title{Two-loop Calculations in the MSSM with FeynArts}
\author{S.~Heinemeyer}
\affiliation{HET, Brookhaven Natl.\ Lab., Upton, New York 11973, USA}

\begin{abstract}
Recent electroweak \twol\ corrections in the Minimal Supersymmetric
Standard Model (MSSM) are reviewed. They have been obtained with the help of
the programs \fa\ and \tc, making use of the recently completed MSSM
model file for \fa. Short examples of how to use the two codes
together with the analytic result for the \order{\gf^2\mt^4}
corrections to the $\rho$-parameter in the MSSM are presented.
\end{abstract}

\maketitle


\section{Introduction}

Theories based on Supersymmetry (SUSY) are widely considered as the
theoretically most appealing extension of the Standard Model
(SM). The Minimal Supersymmetric Standard Model (MSSM) predicts the
existence of scalar partners 
$\tilde{f}_L, \tilde{f}_R$ to each SM chiral fermion, and spin--1/2
partners to the  
gauge bosons and to the scalar Higgs bosons, where two Higgs doublets
are present in the MSSM. So far, the direct search of 
SUSY particles at present colliders has not been successful. 
One can only set lower bounds of ${\cal O}(100 \gev)$ on 
their masses, see \citere{pdg}. 
An alternative way to probe SUSY is to search for the virtual effects of the 
additional particles. The experimental precision has to be matched
with high precision theoretical predictions for the various precision
observables. The most prominent role in this respect is played by the 
$\rho$-parameter, see \citere{rho}:
\BE
 \rho = \frac{1}{1 - \De\rho}, \quad 
 \De\rho = \frac{\Si_Z(0)}{\MZ^2} - \frac{\Si_W(0)}{\MW^2} .
\label{def:rho}
\EE
The radiative corrections to the vector boson
self-energies at zero momentum transfer, $\Si_{Z,W}(0)$, 
constitute the leading, process independent corrections
to many electroweak precision observables, such as the
$W$~boson mass, $\MW$, where
\BE
\de\MW \approx \MW/2\;\;\; \cw^2/(\cw^2 - \sw^2)\; \De\rho ,
\label{delMW}
\EE
with $\cw^2 = 1 - \sw^2 = \MW^2/\MZ^2$.
Within the MSSM the corrections have so far been restricted to
$\oaas$, see \citere{dr2lalals}. 
In order to match the accuracy obtained in the SM and the
(prospective) experimental uncertainties, the leading \twol\ corrections to 
$\De\rho$ at \order{\gf^2\mt^4} are desirable. These corrections are the
lowest order contributions involving non-SM particles with a
$\mt^4$~dependence. 
In the limit of a large SUSY scale, $\msusy \gg \MZ$, the SUSY
particles decouple from the observables, leaving the two Higgs
doublets of the MSSM active. First results for this contribution have
recently been calculated in \citeres{dr2lal2A,dr2lal2B}.

Feynman-diagrammatic \twol\ calculations involve a large number of
diagrams. In the MSSM the additional problem of a proliferation of
scales is apparent. Therefore, in order to perform the calculation as
presented in \citere{dr2lal2A,dr2lal2B}, 
the use of computer algebra programs is
inevitable. In particular, we made use of the amplitude generator
\fa, see \citere{feynarts}, where the MSSM model file has recently been
completed. The reduction of the amplitudes to scalar integrals have
been performed with the program \tc, see \citere{twocalc}.


\section{Techniques for \twol\ calculations}

\begin{figure}[ht!]
\resizebox{\columnwidth}{!}{
\psfig{figure=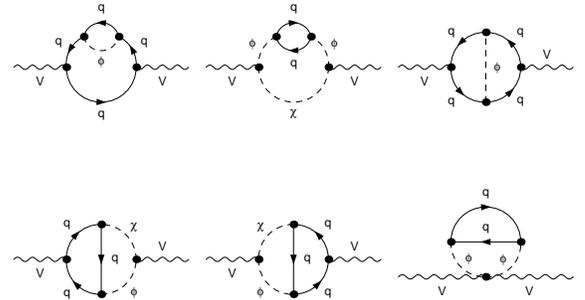,width=4cm,bbllx=070pt,bblly=330pt,
                                         bburx=550pt,bbury=580pt}}

\caption{
Generic diagrams for \order{\gf^2\mt^4} corrections to $\De\rho$.
($V = Z, W$, $f = t, b$, $\phi, \chi = h, H, A, H^\pm, G, G^\pm$)
}
\label{fig:fdeb2l}
\end{figure}

In order to calculate the \order{\gf^2\mt^4} corrections to the
$\rho$-parameter, the diagrams in \reffi{fig:fdeb2l} have to be
evaluated. 

The diagrams and the corresponding amplitudes have been generated with
the help of the \mma\ program \fa\ and the recently accomplished MSSM
model file. This model file contains all relevant information about
the MSSM particles and vertices. (Only MSSM counter term vertices are
missing at present.) \fa\ has been checked in several ways to insure its
reliability. Apart from many self-energy calculations, whole processes
like 
$e^+e^- \to t\bar t$, $e^+e^- \to H^+H^-$, $q\bar q \to t\bar t$ and
$e^+e^- \to \tilde\chi^{0,+} \tilde\chi^{0,-}$ have been calculated by
hand and with \fa. Perfect agreement has been found for all
processes, see~\citeres{feynarts,hep-processes}. 

How easy the evaluation of the amplitudes has become
with \fa\ is demonstrated in the following sequence from a \mma\
session where 
the $Z$~boson self-energy amplitude, corresponding to
\reffi{fig:fdeb2l}, has is obtained. (A detailed guide can be found
in \citere{feynarts}.)

\noindent
\setlength{\unitlength}{1cm}
\begin{picture}(8,0.2)
\thicklines
\put(0,0){\line(1,0){8}}
\end{picture}
(start of \mma\ session)\\
{\tt
$>$
$<<$FeynArts.m;}\\
\phantom{$>$} ($\to$ loading \fa\ into \mma)\\[.5em]
{\tt
$>$
se2 = CreateTopologies[2, 1->1,\\
\phantom{$>$ se2 =} ExcludeTopologies->{Internal}];}\\
\phantom{$>$} ($\to$ \twol\ topologies are created)\\[.5em]
{\tt
$>$
V2V2 = InsertFields[se2, \\
\phantom{$>$ V2V2 =}
               V[2] -> V[2], \\
\phantom{$>$ V2V2 =}
               Model->"MSSM", \\
\phantom{$>$ V2V2 =}
               ExcludeParticles -> \{\dots\} ];}\\
\phantom{$>$} ($\to$ fields are inserted into the topologies, incoming
field {\tt V[2]} = $Z$ and outgoing field {\tt V[2]} are specified,
the model is chosen to be the MSSM)\\[.5em]
{\tt
$>$
Paint[V2V2];}\\
\phantom{$>$} ($\to$ the diagrams are painted (optional), 
see \reffi{fig:fdeb2l})\\[.5em]
{\tt
$>$
V2V2A = CreateFeynAmp[V2V2];}\\
\phantom{$>$} ($\to$ Feynman diagrams are converted into amplitudes)\\[.5em]
{\tt
$>$
V2V2A $>>$ V2V2.amp;} \\
\phantom{$>$} ($\to$ result for the \twol\ amplitude of the $Z$
self-energy is saved in the file {\tt V2V2.amp})\\
(end of \mma\ session) \\
\setlength{\unitlength}{1cm}
\begin{picture}(8,0.2)
\thicklines
\put(0,0.2){\line(1,0){8}}
\end{picture}

The further evaluation of the amplitudes has been performed with the
\mma\ program \tc, see \citere{twocalc}. It performs the reduction of the
self-energy amplitude at the \twol\ level to a basic set of scalar
integrals. The application of \tc\ is demonstrated
in the following sequence from a \mma\ session where
the $Z$~boson self-energy amplitude is processed.

\noindent
\setlength{\unitlength}{1cm}
\begin{picture}(8,0.2)
\thicklines
\put(0,0){\line(1,0){8}}
\end{picture}
(start of \mma\ session)\\
{\tt
$>$
$<<$TwoCalc.m;}\\
\phantom{$>$} ($\to$ loading \tc\ into \mma)\\[.5em]
{\tt
$>$
amp =$<<$V2V2.amp;}\\
\phantom{$>$} ($\to$ the amplitude obtained with \fa\ is loaded)\\[.5em]
{\tt
$>$
SetOptions[ TwoLoop,\\
\phantom{$>$ SetOptions~} CollectFunction -> 0];}\\
\phantom{$>$} ($\to$ options are set.
 {\tt CollectFunction} allows to choose between DREG and DRED)\\[.5em]
{\tt
$>$
res = TwoLoopSum[amp,\\
\phantom{$>$ res =} SelfEnergyPart -> 1];}\\
\phantom{$>$} ($\to$ the amplitude is reduced to scalar integrals)\\[.5em]
{\tt
$>$
res = Collect[res, \\
\phantom{$>$ res =} \{A0[\_], T[\_\_]\}, Simplify];}\\
\phantom{$>$} ($\to$ the result is simplified (optional))\\[.5em]
{\tt
$>$
res $>>$ V2V2.res;}\\
\phantom{$>$} ($\to$ the result for the $Z$~boson self-energy
amplitude in terms of scalar integrals is saved in {\tt V2V2.res})\\
(end of \mma\ session) \\
\setlength{\unitlength}{1cm}
\begin{picture}(8,0.2)
\thicklines
\put(0,0.2){\line(1,0){8}}
\end{picture}

The analytically obtained result now consists of scalar integrals at
the one- and at the \twol\ level. In the case presented here, since
the external momentum is set to zero, these are the functions 
$\rm{A}_0$ and $\rm{T}_{134}$ at the one- and \twol\ level, respectively, 
see \citeres{a0b0c0d0,t134}.

The pure \twol\ diagrams have to be supplemented with the
corresponding \onel\ diagrams with counter term insertion.
In order to obtain the \order{\gf^2\mt^4} corrections an expansion of
the amplitudes up to \order{\mt^4/\MW^4} had to be performed. All
remaining $\MW = \MZ \cw$ have been set to zero.
Furthermore we had to apply the MSSM sum rules for Higgs boson masses,
especially implying for the case $\MZ, \MW \to 0$ that the lightest
MSSM Higgs boson has the mass $\mh = 0$ at tree level.
After adding up all contribution, the expansion of
the result in terms of $\de = (4 - D)/2$ leads to a finite result in
the limit $\de \to 0$.
All relevant details can be found in \citeres{dr2lal2A,dr2lal2B}.


\section{Results for $\De\rho$ and $\MW$}

After extracting the prefactor of $\mt^4/\MW^4$ and setting $\MW$ to
zero, besides $\sbe \equiv \tb/\sqrt{1 + \TQb}$ 
only two mass scales remain: the top quark mass, $\mt$, and the mass
of the $\cp$-odd Higgs boson, $\MA$. The result for $\De\rho$ can be
conveniently expressed in terms of $a \equiv \mt^2/\MA^2$:
\newcommand{\sqa}{\sqrt{1 - 4\,a}}
\def\Liz#1{{\rm Li}_2 \KL #1 \KR}
\BEA
\lefteqn{
\De\rho_1^{\SU} = 3 \, \frac{\gf^2}{128 \,\pi^4} \, \mt^4 \, 
                      \frac{1 - \sbe^2}{\sbe^2 \, a^2} \times \non } \\
&& \Bigg\{ 
   \Liz{ \KL 1 - \sqa \KR/2} \frac{8}{\sqa} \La \non \\
&& - 2\, \Liz{1 - \ed{a}} \KKL 5 - 14 a + 6 a^2 \KKR \non \\
&& + \log^2(a) \KKL 1 + \frac{2}{\sqa} \La \KKR \;
   - \log(a) \Big[ 2 - 20 \, a \Big] \non \\
&& - \log^2 \KL \frac{1 - \sqa}{2} \KR \frac{4}{\sqa} \La \non \\ 
&& + \log \KL \frac{1 - \sqa}{1 + \sqa} \KR 
     \sqa (1 - 2 \, a) \non \\
&& - \log\KL |1/a - 1| \KR \, (a - 1)^2 \non \\
&& + \pi^2 \KKL \frac{2\sqa}{-3 + 12 \, a} \La
                + \ed{3} - 2\, a^2 \frac{\sbe^2}{1 - \sbe^2} \KKR \non \\
&& - 17 a + 19 \frac{a^2}{1 - \sbe^2}
   \Bigg\}
\label{drallma}
\EEA
with 
$
\La = 3 - 13\,a + 11\,a^2 
$ .
As a consistency check, in the limit of $\MA \to \infty$, $a \to 0$,
we obtain
\BE
\De\rho_1^{\SU} = 3 \, \frac{\gf^2}{128 \,\pi^4} \, \mt^4 
                  \KKL 19 - 2 \pi^2 \KKR . 
\EE
This is the SM result with $\MH \to 0$. It shows that the MSSM
decouples to the SM limit (also at the \twol\ level) 
when the new scales, here $\MA$, is made large.

In the limit of large $\tb$ (i.e.\ $(1 - \sbe^2) \ll 1$) one obtains
\BE
\De\rho_1^{\SU} = 3 \, \frac{\gf^2}{128 \,\pi^4} \, \mt^4 \, 
\KKL \frac{19}{\sbe^2} - 2\,\pi^2 + {\cal O}(1 - \sbe^2) \KKR .
\label{drallmalargetb}
\EE
Thus for large $\tb$ the SM limit with $\MH \to 0$ is reached.

The numerical result obtained for the \order{\gf^2\mt^4} corrections
to $\De\rho$ vary around $-2 \times 10^{-5}$ and are of the same size
(but with opposite sign) as the leading \order{\al\als} MSSM
corrections originating from the scalar top and bottom
sector, see \citere{dr2lalals}, if the common soft SUSY-breaking scale is
chosen to be $\msusy \approx 500 \gev$. For a larger value, 
$\msusy \approx 1000 \gev$, the new \order{\gf^2\mt^4} corrections are
about three times larger than the \order{\al\als} contributions.
Furthermore 
it is well known that the \order{\gf^2\mt^4} SM result with $\MH^{\SM} = 0$
underestimates the result with $\MH^{\SM} \neq 0$ by one order of
magnitude. One can expect (see \citere{dr2lal2A}) a similar effect in
the MSSM once higher 
order corrections to the Higgs boson sector are properly taken into
account, which can enhance $\mh$ up to 
$\mh \lsim 130 \gev$, see \citere{mhiggs2l}.

\begin{figure}[ht!]
\resizebox{\columnwidth}{!}{
\epsfig{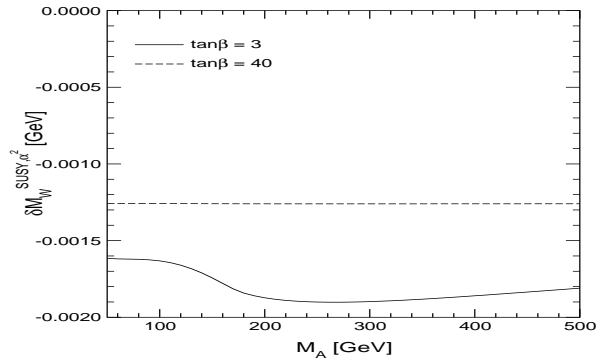}}

\caption{
The induced shift in $\MW$ is shown as a function of $\MA$ for 
$\tb = 3, 40$.
}
\label{fig:delMW}
\end{figure}

With the help of \refeq{delMW} the shift in the $W$~boson mass can be
evaluated. In \reffi{fig:delMW} the induced shift for 
$\MW$ is shown as a function of $\MA$ for $\tb = 3, 40$. 
The effect on $\de\MW$ varies between $-1.5 \mev$ and $-2 \mev$ 
for small $\tb$ and
is almost constant, $\de\MW = -1.25 \mev$, for $\tb = 40$.
The constant behavior can be
explained by the analytical decoupling of $\tb$ in \refeq{drallma}
when $\tb \gg 1$, see \refeq{drallmalargetb}.

\vspace{-1.5em}
\section{Conclusions}

We presented the calculation of the SUSY contributions of
\order{\gf^2\mt^4} to the $\rho$-parameter. In order to obtain these
\twol\ corrections, the computer algebra programs \fa\ and \tc\ have
been used. Examples of the handling of these programs were given. A
compact analytical result in terms of $\mt^2/\MA^2$ has been
derived. The induced shift in $\MW$ varies between $-1 \mev$ and 
$-2 \mev$, depending on $\MA$ and $\tb$.


\section{Acknowledgments}

S.H. thanks T.~Hahn, C.~Schappacher and other members of the TP,
Universit\"at Karlsruhe, Germany, for their effort put into \fa\ and
the new MSSM model file. S.H. gratefully acknowledges the
collaboration with G.~Weiglein, with whom the results presented here
have been obtained.


\begin{thebibliography}{1}

\bibitem{pdg} Part. Data Group,
              {\em Eur.\ Phys.\ Jour.\ }{\bf C15} (2000) 1.

\bibitem{rho} M.~Veltman, 
              {\em Nucl. Phys.} {\bf B123} (1977) 89. 

\bibitem{dr2lalals} A.~Djouadi, P.~Gambino, S.~Heinemeyer, W.~Hollik,
                    C.~J\"unger and G.~Weiglein, 
                    {\em Phys. Rev. Lett.} {\bf 78} (1997) 3626; 
                    {\em Phys. Rev.} {\bf D57} (1998) 4179.

\bibitem{dr2lal2A} S.~Heinemeyer and G.~Weiglein,
                   {\em in preparation}.

\bibitem{dr2lal2B} S.~Heinemeyer and G.~Weiglein, to appear in the 
                   proceedings of the RADCOR2000, Carmel, Sep.~2000,
                   hep-ph/0102317.

\bibitem{feynarts} T.~Hahn, 
                   hep-ph/0012260;
                   (see {\tt www.feynarts.de}).

\bibitem{twocalc} G.~Weiglein, R.~Scharf and M.~B\"ohm,
                  {\em Nucl. Phys.} {\bf B416} (1994) 606.

\bibitem{hep-processes} see {\tt www.hep-processes.de} .

\bibitem{a0b0c0d0} G.~Passarino and M.~Veltman, 
                   {\em Nucl. Phys.} {\bf B160} (1979) 151.

\bibitem{t134} A.~Davydychev and J.~Tausk, 
               {\em Nucl. Phys.} {\bf B397} (1993) 123;
               F.~Berends and J.~Tausk, 
               {\em Nucl. Phys.} {\bf B421} (1994) 606.

\bibitem{mhiggs2l} S.~Heinemeyer, W.~Hollik and G.~Weiglein,
                    {\em Phys. Rev.} {\bf D58} (1998) 091701;
                    {\em Phys. Lett.} {\bf B440} (1998) 296;
                    {\em Eur. Phys. Jour.} {\bf C9} (1999) 343.



\end{thebibliography}


\end{document}